\documentclass[prc,showpacs,nofootinbib,preprint]{revtex4-1}
\usepackage{epsfig,graphics,color}
\usepackage{subfigure}
\usepackage{bbm}
\usepackage{amssymb}
\usepackage{amsmath}

\newcommand{\bra}[1]{\langle #1|}
\newcommand{\ket}[1]{|#1\rangle}

\newcommand{\roundbra}[1]{(#1|}
\newcommand{\roundket}[1]{|#1)}

\def\one{{\bf 1}\,}

\def\tr{{\rm tr} \,}

\def\vslash{v \hspace{-1.7mm}/}

\def\w2{\tilde w^2}
\def\ws2{1}

\newcommand{\smallfrac}[2]{\textstyle{\frac #1 #2}}

\begin{document}
\title{ Combined heavy-quark symmetry and large-$N_c$ operator analysis for 2-body counterterms in the chiral Lagrangian with\\ $D$ mesons and charmed baryons}
\author{Daris Samart$^{1,2,3}$}\email{daris.sa@rmuti.ac.th : corresponding author}
\author{Chakrit Nualchimplee$^2$} \author{Yupeng Yan$^{1,3}$}
\affiliation{$^1$ School of Physics, Suranaree University of Technology, Nakhon Ratchasima, 30000, Thailand}
\affiliation{$^2$ Department of Applied Physics, Rajamangala University of Technology Isan, Nakhon Ratchasima, 30000, Thailand}
\affiliation{$^3$ Thailand Center of Excellence in Physics (ThEP), Commission on Higher Education, Bangkok 10400, Thailand}
\date{\today}
\begin{abstract}
We construct, in the work, chiral $SU(3)$ Lagrangian with
$D$ mesons of spin $J^P=0^-$ and $J^P=1^-$ and charmed baryons of spin $J^P=1/2^+$ and $J^P=3/2^+$. There are 42 leading two-body counter-terms involving two charmed baryon fields and two $D$ meson fields in the constructed Lagrangian. The heavy-quark
spin symmetry leads to 35 sum rules while the large-$N_c$ operator analysis predicts 29 ones at the next-to leading order of $1/N_c$ expansion. The combination of the sum rules from both the heavy-quark symmetry and the large-$N_c$ analysis results in 38 independent sum rules which reduces the number of free parameters in the chiral Lagrangian down to 4 only. This is a remarkable result demonstrating the consistency of the heavy-quark symmetry and large-$N_c$ operator analysis.
\end{abstract}

\maketitle

\section{Introduction}
The chiral $SU(3)$ Lagrangian has been extensively applied for studying meson and baryon resonances in the charm sector. For instance, the leading order chiral Lagrangian is used to study s-wave scatterings of Goldstone bosons off $D$ mesons \cite{Kolomeitsev-Lutz-2004,Hofmann-Lutz-2004,Lutz-Soyeur-2006,Guo-Hanhart-Krewald-Meissner-2008} and Goldstone bosons with ground state open charmed baryons \cite{Lutz-Kolomeitsev-2004-charm,Lutz:2005ip} based on coupled-channel calculations. A rich spectrum of s-wave and d-wave exotic charmed baryon resonances is dynamically generated by zero-range t-channel vector meson exchange interactions \cite{Hofmann-Lutz-2005,Hofmann-Lutz-2006,Tolos-2004}.

The application of chiral Lagrangian in the hidden charmed baryons is also received a number of attentions by using the couple channel dynamics framework. This study is motivated by FAIR at GSI with \={P}ANDA experiment which will be in a promising position to provide more data on new exotic states of charm and strangeness degree of freedom \cite{Lutz:2009ff}. 
The couple channel calculations with the chiral Lagrangian have been done to explore new resonances in the high energy region. For example, the nucleon like resonances with the hidden charm quantum number are dynamically generated by using chiral Lagrangian with various models such as t-channel vector meson exchange picture with $SU(4)$ flavor-symmetry \cite{Hofmann-Lutz-2005,Lutz:2006ya}, local hidden gauge formalism \cite{Wu:2010jy,Wu:2010vk}, Weinberg-Tomozawa interaction with heavy-quark symmetry \cite{Garcia-Recio:2013gaa} and a combination of heavy-quark and local hidden gauge symmetries \cite{Xiao:2013yca}.
Most of these approaches have employed the heavy-quark symmetry to incorporate the chiral Lagrangian in their works. The main interpretation of heavy-quark symmetry is that in the infinite limit of heavy-quark mass the pseudoscalar and vector mesons with one heavy-quark form a degeneracy state \cite{Isgur:1989vq} as well as for charmed baryons with spin-$\frac12$ and $\frac32$ \cite{Georgi:1990cx}. It is well known that the heavy quark symmetry plays an important role in the heavy quark sector. However, deviations of the heavy-quark symmetry is expected because the charm quark mass is not extremely larger than a typical confinement scale $\Lambda_{\rm QCD}$ \cite{Flynn:1992fm}. Therefore, in order to relate more closely to QCD, one might take an additional approximate symmetry of QCD into account.

Large-$N_c$ QCD, on the other hand, is another approximate symmetry of QCD. This idea is originally suggested by Ref. \cite{'tHooft:1973jz} which notes that one can consider the color number of degrees of freedom ($N_c$) to be large and expand it in power of $1/N_c$. By using such expansion, a number of simplifications of QCD occurs in the large number of colors limit. This approach has been shown very useful in the study of baryons at the low-energy regime (for review see \cite{Jenkins:1998wy,Matagne:2014lla}). An interesting prediction of large-$N_c$ QCD is the approximate degeneracy of the baryon octet and decuplet states \cite{Witten:1979kh} forming a super multiplet in the large-$N_c$ limit. According to $N_c$ counting rules for meson-baryon systems \cite{Witten:1979kh}, subsequently, this leads to a spin-flavor symmetry for baryons at the large-$N_c$ limit \cite{Gervais:1983wq,Dashen:1993as}. In particular, the combination of the heavy-quark symmetry and large-$N_c$ analysis provides predictions of heavy baryons of excited and exotic states \cite{Lee:1998pq,Chow:1999hm,Lee:2000wb,AzizaBaccouche:2001pu,Wessling:2004ag,Cohen:2005bx,Semay:2008wn}.

In order to perform a complete couple channel calculation, one needs to take both the short-rang and long-range interactions into account. The short range forces between $D$ mesons and charmed baryons may be parameterized in terms of contact interactions.  In a previous work \cite{Lutz:2011fe}, those counter terms were analysed systematically by using the heavy-quark symmetry and large-$N_c$ sum rules. The 26 counter terms were correlated so that only 7 unknown parameters remain. Furthermore, a corresponding analysis for describing the short-range interaction of the Goldstone bosons with the charmed baryons has been performed by Ref. \cite{Lutz:2014jja}. The 36 counter terms were reduced down to 6 unknown parameters only.

This work is to prepare for systematic coupled-channel computations on resonances of hidden-charm
baryon (charm quantum number equals to zero for example anti-$D$ mesons and charmed baryons interactions) and doubly charmed baryon (charm quantum number equals to two e.g. $D$ mesons with charmed baryons system) sectors. A systematic construction of the leading 2-body counter terms for the $D$ mesons with $J^P=0^-$ and $1^-$\, and the charmed baryons with $J^P=\frac12^+$ and $\frac32^+$\, quantum numbers is considered. The low-energy constants of the chiral Lagrangian are correlated by using heavy-quark symmetry with a suitable super multiplets of $D$ mesons \cite{Lutz-Soyeur-2006,Georgi:1990um,Wise:1992hn,Casalbuoni:1996pg,Mehen:2004uj} and charmed baryons \cite{Georgi:1990cx,Yan:1992gz,Cho:1992gg}. To work out the sum rules from large-$N_c$ scheme, we study baryon matrix elements of the correlation functions \cite{Lutz:2011fe,Lutz:2014jja,Krause:1990xc,Lutz:2010se}. The technology of large-$N_c$ operator analysis for light-quark baryons has been developed by Refs. \cite{Lutz:2014jja,Lutz:2010se,Luty:1993fu,Dashen:1993jt,Dashen:1994qi,Jenkins:1996de} and shown well applicable to charmed baryons \cite{Lutz:2014jja}.

This paper is organized as follows. The chiral Lagrangian with $D$ mesons and charmed baryons are constructed in Section II.
The implications of the heavy-quark spin symmetry on the coupling constants are worked out in Section III while the Large-$N_c$ operator analysis of charmed baryon matrix elements is performed in Section IV.  Finally, Section V is devoted to a summary of main results.

\section{Chiral Lagrangian with $D$ mesons and charmed baryons} \label{section:chiral-lagrangian}
This section is devoted to construct the chiral Lagrangian of the $D$ mesons with the charmed baryons. The construction rules for the chiral $SU(3)$ Lagrangian density are referred to Refs. \cite{Krause:1990xc,GL84,Ecker89,Borasoy,Birse,Becher,Fuchs} for more technical details. In this work, we follow the conventions and notations from Refs. \cite{Lutz:2010se,Lutz:2011fe,Lutz:2014jja}.

We focus on the residual short-range interactions described by local two-body counter terms. The basic building blocks of the chiral Lagrangian in this study are
\begin{eqnarray}
D\,,\quad D^{\mu\nu}\,,\quad B_{[\bar 3]}\,,\quad B_{[6]}\,,\quad B_{[6]}^\mu\,,
\label{def-fields}
\end{eqnarray}
where the anti-triplet pseudoscalar meson fields  $D(J^P\!\!=\!0^-)$, vector meson fields $D^{\mu \nu}(J^P\!\!=\!1^-)$,
charmed baryon anti-triplet fields $B_{[\bar3]}(J^P\!\!=\!{\textstyle{1\over2}}^+)$ and charmed baryon sextet fields $B_{[6]}(J^P\!\!=\!{\textstyle{1\over2}}^+)$, $B_{[6]}^\mu(J^P\!\!=\!{\textstyle{3\over2}}^+)$ are heavy fields in ChPT \cite{Krause:1990xc}. Their $SU(3)$ multiplet forms and transformation properties under the chiral $SU(3)$ rotation are referred to Refs. \cite{Lutz:2010se,Lutz:2011fe,Lutz:2014jja} in more details. The leading order two-body counter terms involving two $D$ meson and two charmed baryon fields can be discriminated into terms with two spin-$\frac12$ charmed baryon fields, ${\mathcal L}^{(c)}$, two spin-$\frac32$ charmed baryon fields, ${\mathcal L}^{(d)}$, and the mixing spin-$\frac12$ and $\frac32$ fields, ${\mathcal L}^{(e)}$,
\begin{eqnarray}
{\mathcal L}^{\rm counter} = {\mathcal L}^{(c)} + {\mathcal L}^{(d)} + {\mathcal L}^{(e)}\,.
\label{counter-terms}
\end{eqnarray}
The chiral Lagrangians, ${\mathcal L}^{(c)}$, ${\mathcal L}^{(d)}$ and ${\mathcal L}^{(e)}$ are used to describe low-energy s-wave scattering and hence we consider the chiral power order $Q^0$ only.  A detailed discussion of the spin structures of two-body systems and flavor $SU(3)$ invariance of the Lagrangian in Eqs. (\ref{counter-terms}) can be found in Refs. \cite{Lutz:2010se} and \cite{Lutz:2014jja}.

We begin with two charmed baryon spin-$\frac12$ fields, and get 20 leading order terms in ${\mathcal L}^{(c)}$,
\allowdisplaybreaks
\begin{eqnarray}\label{def-Lc}
&& {\mathcal L}^{(c)} = D\,\Big\{c_{1,[\bar3\bar3]}^{(S)}  \,\bar B_{[\bar3]}\,B_{[\bar3]} + \textstyle{\frac12}\,c_{2,[\bar3\bar3]}^{(S)}\, {\rm tr}\,(\bar B_{[\bar3]}\,B_{[\bar3]})\Big\}\,\bar D
\nonumber\\
&& \qquad +\, D\,\Big\{c_{1,[66]}^{(S)}  \,\bar B_{[6]}\,B_{[6]} + \textstyle{\frac12}\,c_{2,[66]}^{(S)}\, {\rm tr}\,(\bar B_{[6]}\,B_{[6]}) \Big\}\,\bar D
\nonumber\\
&& \qquad +\, D\,c_{1,[\bar36]}^{(S)}  \,\Big\{\bar B_{[6]}\,B_{[\bar3]} + {\rm h.c.} \Big\}\,\bar D
\nonumber \\
&& \qquad -\, \frac{1}{2}\,D_{\mu \nu}\,\Big\{\tilde c_{1,[\bar3\bar3]}^{(S)} \,\bar B_{[\bar3]}\,B_{[\bar3]}
+ \textstyle{\frac12}\,\tilde c_{2,[\bar3\bar3]}^{(S)}\, {\rm tr}\,(\,\bar B_{[\bar3]}\,B_{[\bar3]})
\Big\}\,\bar D^{\mu\nu}
\nonumber\\
&& \qquad -\, \frac{1}{2}\, D_{\mu\nu}\,\Big\{\tilde c_{1,[66]}^{(S)}  \,\bar B_{[6]}\,B_{[6]}
+ \textstyle{\frac12}\,\tilde c_{2,[66]}^{(S)}\, {\rm tr}\,(\bar B_{[6]}\,B_{[6]})\Big\}\,\bar D^{\mu\nu}
\nonumber \\
&& \qquad -\, \frac{1}{2}\,D_{\mu \nu}\,\tilde c_{1,[\bar36]}^{(S)}  \,\Big\{\bar B_{[6]}\,B_{[\bar3]} + {\rm h.c.} \Big\}\,\bar D^{\mu\nu}
\nonumber \\
&& \qquad + \, \frac{i}{M_c}\,D_{\mu \nu}\,\Big\{
c_{1,[\bar3\bar3]}^{(A)}\,\bar B_{[\bar3]}\, \gamma^\mu\,\gamma_5\,B_{[\bar3]}
+ \textstyle{\frac12}\,c_{2,[\bar3\bar3]}^{(A)}\,{\rm tr} \,(\bar B_{[\bar3]}\, \gamma^\mu\,\gamma_5\,B_{[\bar3]})\Big\} \,(\partial^\nu \bar D)
+ {\rm h.c.}
\nonumber \\
&& \qquad + \, \frac{i}{M_c}\,D_{\mu \nu}\,\Big\{
c_{1,[66]}^{(A)}\,\bar B_{[6]}\, \gamma^\mu\,\gamma_5\,B_{[6]}
+ \textstyle{\frac12}\,c_{2,[66]}^{(A)}\,{\rm tr} \,(\bar B_{[6]}\, \gamma^\mu\,\gamma_5\,B_{[6]})\Big\} \,(\partial^\nu \bar D)
+ {\rm h.c.}
\nonumber \\
&& \qquad - \, \frac{i}{2\,M_c}\,c_{1,[\bar36]}^{(A)}\,\Big\{
D_{\mu \nu}\,\bar B_{[6]}\, \gamma^\mu\gamma_5\,\,B_{[\bar3]}\,(\partial^\nu \bar D)
- (\partial^\nu D)\,\bar B_{[6]}\, \gamma^\mu\gamma_5\,\,B_{[\bar3]}\,\bar D_{\mu \nu}\, \Big\}
+ {\rm h.c.}
\nonumber\\
&& \qquad + \,\frac{1}{4 \,M_c}\, \epsilon^{\mu \nu \alpha \beta}
D_{\mu \nu} \Big\{
\tilde c_{1,[\bar3\bar3]}^{(A)}\,\bar B_{[\bar3]} \, \gamma_\alpha \,\gamma_5\,B_{[\bar3]}
+ \textstyle{\frac12}\,\tilde c_{2,[\bar3\bar3]}^{(A)}\,{\rm tr} \,(\bar B_{[\bar3]} \, \gamma_\alpha\,\gamma_5 \,B_{[\bar3]})\Big\}
(\partial^\tau \bar D_{\tau \beta} ) + {\rm h.c.}
\nonumber\\
&& \qquad + \,\frac{1}{4 \,M_c}\, \epsilon^{\mu \nu \alpha \beta}
D_{\mu \nu} \Big\{
\tilde c_{1,[66]}^{(A)}\,\bar B_{[6]}\, \gamma_\alpha\,\gamma_5 \,B_{[6]}
+ \textstyle{\frac12}\,\tilde c_{2,[66]}^{(A)}\,{\rm tr} \,(\bar B_{[6]} \, \gamma_\alpha\,\gamma_5 \,B_{[6]})\Big\}
(\partial^\tau \bar D_{\tau \beta} ) + {\rm h.c.}
\\
&& \qquad - \,\frac{1}{4 \,M_c}\,\tilde c_{1,[\bar36]}^{(A)}\, \epsilon^{\mu \nu \alpha \beta}
\Big\{D_{\mu \nu} \,\bar B_{[6]}\, \gamma_\alpha\,\gamma_5 \,B_{[\bar3]}
(\partial^\tau \bar D_{\tau \beta} )
+ (\partial^\tau D_{\tau \beta} )\,\bar B_{[6]} \, \gamma_\alpha\,\gamma_5 \,B_{[\bar3 ]}
\,\bar D_{\mu\nu} \Big\} + {\rm h.c.}\,,\nonumber
\end{eqnarray}
where $\bar D \equiv D^\dagger$ and $M_c$ is the mass of the charm quark, ``\,${\rm tr}$\," stands for the trace over SU(3) flavor space.
There are 10 terms at leading order terms involving two baryon spin-$\frac32$ fields, that is,
\begin{eqnarray}
&& {\mathcal L}^{(d)} = - D\,\Big\{d_{1,[66]}^{(S)} \,\bar B_{[6]}^\alpha\,g_{\alpha\beta} \, B_{[6]}^\beta
+ \textstyle{\frac12}\,d_{2,[66]}^{(S)}\, {\rm tr}\,(\bar B_{[6]}^\alpha\,g_{\alpha\beta} \, B_{[6]}^\beta) \Big\}\,\bar D
\nonumber\\
&& \qquad +\, \frac{1}{2}\,D_{\mu \nu}\,\Big\{\tilde d_{1,[66]}^{(S)} \,\bar B_{[6]}^\alpha\,g_{\alpha\beta} \, B_{[6]}^\beta
+ \textstyle{\frac12}\,\tilde d_{2,[66]}^{(S)}\, {\rm tr}\,(\bar B_{[6]}^\alpha\,g_{\alpha\beta} \, B_{[6]}^\beta) \Big\}
\,\bar D^{\mu\nu}
\nonumber\\
&& \qquad + \, \frac{i}{4 }\,\epsilon_{\mu \nu \alpha \beta }\,D^{\mu \nu}\,\Big\{
d_{1,[66]}^{(E)}\,\bar B_{[6]}^\alpha\, B_{[6]}^\beta
+ \textstyle{\frac12}\,d_{2,[66]}^{(E)}\,{\rm tr} \,(\bar B_{[6]}^\alpha\, B_{[6]}^\beta)\Big\} \,\bar D
 + {\rm h.c.}
\nonumber\\
&& \qquad + \, \frac{1}{2}\,D_{\beta \mu}\,\Big\{\tilde d_{1,[66]}^{(E)} \,\bar B_{[6]}^{\tau}\,g_{\tau\alpha}\, B_{[6]}^\beta
+ \textstyle{\frac12}\,\tilde d_{2,[66]}^{(E)}\, {\rm tr}\,(\bar B_{[6]}^{\tau}\,g_{\tau\alpha}\, B_{[6]}^\beta) \Big\}
\,\bar D^{\alpha \mu}\nonumber\\
&& \qquad - \, \frac{1}{2}\,D_{\alpha \mu}\,\Big\{\tilde d_{3,[66]}^{(E)} \,\bar B_{[6]}^{\alpha}\,g_{\beta\tau}\, B_{[6]}^\tau
+ \textstyle{\frac12}\,\tilde d_{4,[66]}^{(E)}\, {\rm tr}\,(\bar B_{[6]}^{\alpha}\,g_{\beta\tau}\, B_{[6]}^\tau) \Big\}
\,\bar D^{\beta \mu}\,.
\label{def-Ld}
\end{eqnarray}
We find the following 12 terms involving spin-$\frac12$ and -$\frac32$ charmed baryons,
\begin{eqnarray}
{\mathcal L}^{(e)} &=& \frac{i}{4}\,\epsilon_{\mu \nu \alpha \beta}\,\Big\{
e^{(A)}_{1,[66]}\,\,D^{\alpha \beta} \,
\bar B_{[6]}^\mu \,\gamma^{\nu}\, \gamma_5\,B_{[6]} \, \bar D
+ e^{(A)}_{2,[66]} D\,\bar B_{[6]}^\mu \,\gamma^{\nu}\,\gamma_5\,B_{[6]} \, \bar D^{\alpha \beta} \Big\} + {\rm h.c.}
\nonumber\\
&+& \frac{i}{8}\,\epsilon_{\mu \nu \alpha \beta}\Big\{
e^{(A)}_{3,[66]}\, \,D^{\alpha \beta} \,
{\rm tr}\,(\bar B_{[6]}^\mu \,\gamma^{\nu}\, \gamma_5\,B_{[6]}) \, \bar D
+ e^{(A)}_{4,[66]}D\,{\rm tr}\,(\bar B_{[6]}^\mu \,\gamma^{\nu}\,\gamma_5\,B_{[6]}) \, \bar D^{\alpha \beta} \Big\} + {\rm h.c.}
\nonumber\\
&+& \frac{i}{4}\,\epsilon_{\mu \nu \alpha \beta}\Big\{
e^{(A)}_{1,[\bar36]}\, D^{\alpha \beta} \,
\bar B_{[6]}^\mu \,\gamma^{\nu}\, \gamma_5\,B_{[\bar3]} \, \bar D
+ e^{(A)}_{2,[\bar36]}\,D\,\bar B_{[6]}^\mu \,\gamma^{\nu}\,\gamma_5\,B_{[\bar3]} \, \bar D^{\alpha \beta} \Big\} + {\rm h.c.}
\nonumber\\
&+& \frac{1}{2}\,\Big\{ \tilde  e^{(A)}_{1,[66]}\,  D_{\alpha \nu} \,
\bar B_{[6]}^\mu \, \gamma^{\nu}\,\gamma_5\,B_{[6]}\, g^{\alpha\beta}\,\bar D_{\beta\mu}
- \tilde  e^{(A)}_{2,[66]}\,  D_{\alpha \mu} \,
\bar B_{[6]}^\mu \, \gamma^{\nu}\,\gamma_5\,B_{[6]} \,g^{\alpha\beta}\,\bar D_{\beta \nu} \Big\}  + {\rm h.c.}
\nonumber\\
&+& \frac{1}{4}\,\Big\{\tilde  e^{(A)}_{3,[66]}\,  D_{\alpha \nu} \,
{\rm tr}\,(\bar B_{[6]}^\mu \, \gamma^{\nu}\,\gamma_5\,B_{[6]})\, g^{\alpha\beta}\,\bar D_{\beta\mu}
- \tilde  e^{(A)}_{4,[66]}\,  D_{\alpha \mu} \,
{\rm tr}\,(\bar B_{[6]}^\mu \, \gamma^{\nu}\,\gamma_5\,B_{[6]}) \,g^{\alpha\beta}\,\bar D_{\beta \nu} \Big\}  + {\rm h.c.}
\nonumber\\
&+& \frac{1}{2}\,\Big\{\tilde  e^{(A)}_{1,[\bar36]}\,  D_{\alpha \nu} \,
\bar B_{[6]}^\mu \, \gamma^{\nu}\,\gamma_5\,B_{[\bar3]}\, g^{\alpha\beta}\,\bar D_{\beta\mu}
- \tilde  e^{(A)}_{2,[\bar36]}\,  D_{\alpha \mu} \,
\bar B_{[6]}^\mu \, \gamma^{\nu}\,\gamma_5\,B_{[\bar3]} \,g^{\alpha\beta}\,\bar D_{\beta \nu} \Big\}  + {\rm h.c.}  \,.
\label{def-Le}
\end{eqnarray}
As shown in Eqs. (\ref{def-Lc}), (\ref{def-Ld}) and (\ref{def-Le}), we got totally 20 + 10 + 12 = 42 leading order two-body counter terms. In the following sections, we will use the heavy-quark symmetry and the large $N_c$ operator analysis to correlate the coupling constants introduced here.

\section{Heavy quark mass expansion}\label{section:HQS}

At the limit of infinite charm quark mass, the $D$ mesons with spin-$0^-$ and -$1^-$ are replaced by a spin flip of the charm quark. This is also true for the charmed baryons with spin-$\frac12^+$ and -$\frac32^+$\,. One can imply effects of the heavy-quark symmetry for the chiral Lagrangians in Eqs. (\ref{def-Lc}), (\ref{def-Ld}) and (\ref{def-Le}) by introducing auxiliary and slowly varying fields, $P_\pm(x)$\,, $P^\mu_\pm(x)$\,, $B_\pm(x)$\,, $B^\mu_\pm(x)$ and $B^{[3]}_\pm$. We decompose the $D$ meson and charmed baryons fields into such fields \cite{Lutz-Soyeur-2006,Lutz:2011fe,Lutz:2014jja}
\begin{eqnarray}
&& D(x) \;\;\;\;\,\,\!= e^{-i\,(v\cdot x) \,M_c}\,P_{+}(x) +e^{+i\,(v\cdot x) \,M_c}\,P_{-}(x)\,,
\nonumber\\
&& D^{\mu \nu}(x) \;\;\!=
i\,e^{-i\,(v\cdot x) \,M_c}\,\Big\{v^\mu\,P^\nu_{+}(x)-v^\nu \,P^\mu_{+}(x)
+ \frac{i}{M_c}\,\Big( \partial^\mu P^\nu_+-\partial^\nu P^\mu_+\Big)\Big\}
\nonumber\\
&& \qquad \quad \;\;\,+\,i\,e^{+i\,(v\cdot x) \,M_c}\,\Big\{v^\mu\,P^\nu_{-}(x)-v^\nu \,P^\mu_{-}(x)
- \frac{i}{M_c}\,\Big( \partial^\mu P^\nu_--\partial^\nu P^\mu_-\Big)\Big\}\,,
\nonumber\\
&& B_{[6]}(x) \;\;\,\,\!= e^{-i\,(v\cdot x) \,M^{1/2}_{[6]}}\,B_{+}(x) +e^{+i\,(v\cdot x) \,M^{1/2}_{[6]}}\,B_{-}(x)\,,
\nonumber\\
&& B^\mu_{[6]}(x) \;\;\,\,\!= e^{-i\,(v\cdot x) \,M^{3/2}_{[6]}}\,B^\mu_{+}(x) +e^{+i\,(v\cdot x) \,M^{3/2}_{[6]}}\,B^\mu_{-}(x)\,,
\nonumber\\
&& B_{[\bar 3]}(x) \;\;\,\,\!= e^{-i\,(v\cdot x) \,M^{1/2}_{[\bar 3]}}\,B^{+}_{[\bar 3]}(x) +e^{+i\,(v\cdot x) \,M^{1/2}_{[\bar 3]}}\,B^{-}_{[\bar 3]}(x)\,,
\label{non-relativistic-expansion}
\end{eqnarray}
where the 4-velocity $v$ is normalized with $v^2=1$. The parameters $M^{1/2}_{[6]}$\,, $M^{3/2}_{[6]}$\, and $M^{1/2}_{[\bar3]}$ are the chiral limit masses of the two sextet baryons with spin-$\frac12$ and -$\frac32$ and anti-triplet baryons respectively and, the implications of those parameters in heavy-quark limit can be seen in Ref. \cite{Lutz:2014jja}. The time and spatial derivatives of the fields $\partial_\alpha P_\pm$\,, $\partial_\alpha B_\pm$ are small and can be neglected because of the slow varying of the fields in Eq. (\ref{non-relativistic-expansion}).

We follow the formalism for multiplet fields of $D$ mesons and charmed baryons developed in
\cite{Wise:1992hn,Casalbuoni:1996pg,Mehen:2004uj,Yan:1992gz,Cho:1992gg}, and introduce the multiplet fields $\mathcal{H}$, $H_{[6]}^\mu$ and $H_{[\bar 3]}$ connected to the fields $P_+$ and $P_+^\mu$, and $B_+$ and $B_+^\mu$ respectively as follows
\footnote{Note that
${\tr } \gamma_5\,\gamma_\mu \,\gamma_\nu \,\gamma_\alpha \,\gamma_\beta =
-4\,i\,\epsilon_{\mu \nu \alpha \beta }
$ in the convention used in this work.} :
\begin{eqnarray}
&& \mathcal{H} = \frac{1}{2}\,\Big( 1 + \vslash \Big)\,\Big(\gamma_\mu\,P^\mu_+  +i\, \gamma_5\,P_+ \Big)
\nonumber\\
&& \mathcal{\bar H} = \gamma_0\,\mathcal{H}^\dagger\,\gamma_0 =
\Big(P^\dagger_{+,\mu} \,\gamma^\mu +P^\dagger_+ \,i\, \gamma_5 \Big)\,\frac{1}{2}\,\Big( 1 + \vslash \Big) \,,
\nonumber\\
&& H^\mu_{[6]} = \frac{1}{\sqrt 3}\,(\gamma^\mu + v^\mu)\,\gamma_5\,\frac{1+ \vslash}{2}\,B_{+} + \frac{1+ \vslash}{2}\,B_{+}^\mu\,,\qquad
 \bar H^\mu_{[6]} =  \big(H^\mu_{[6]}\big)^\dagger\,\gamma_0 \,,
 \nonumber\\
&& H_{[\bar 3]} = \frac{1+ \vslash}{2}\,B^{+}_{[\bar 3]}\,, \qquad
\bar H_{[\bar 3]} = \left( H_{[\bar 3]} \right)^\dagger\,\gamma_0 \,
\nonumber\\
&& P^\mu_+ \,v_\mu =0 \, , \qquad  B^\mu_+ \,v_\mu =0 \, , \qquad  v^2=1\,.
\label{phase-convention}
\end{eqnarray}
The transformation properties under $SU_v(2)$ heavy-quark spin and Lorentz symmetries of multiplet fields in Eq.~(\ref{phase-convention}) can be found in the literatures mentioned above. Using the rules demonstrated in Refs. \cite{Lutz:2011fe} and \cite{Lutz:2014jja} for constructing heavy-quark spin invariant effective Lagrangian, it is straightforward to construct the effective Lagrangian bearing the structures detailed in Eq.~(\ref{counter-terms}). The Lagrangian takes the form,
\begin{eqnarray}
{ \mathcal L}^{(H)} &=&\frac{1}{2}\,{\rm Tr }\, \mathcal{H}\,   \Big\{ f^{(S)}_1\,\bar H_{[\bar3]} \, H_{[\bar3]} +  f^{(S)}_2\,{\rm tr }\,\bar H_{[\bar3]} \, H_{[\bar3]}
+ f^{(S)}_3\,\bar H_{[6]}^\mu \,g_{\mu\nu}\, H_{[6]}^\nu
+  f^{(S)}_4\, {\rm tr} \, \bar H_{[6]}^\mu \,g_{\mu\nu}\, H_{[6]}^\nu \Big\}\,\mathcal{\bar H} \nonumber \\
&-& \,\frac{1}{4}\,{\rm Tr } \, \mathcal{H}\, \Big\{ f^{(A)}_1\,\bar H_{[6]}^\mu\,H_{[\bar3]}
+ {\rm h.c.}\Big\}\,\gamma_\mu\,\gamma_5\,\mathcal{\bar H}
\nonumber \\
&+& \,\frac{1}{4}\,{\rm Tr } \,\mathcal{H}\,  \Big\{ f^{(T)}_1\,\bar H_{[6]}^\mu \, H_{[6]}^\nu
+ f^{(T)}_2\,{\rm tr}\, \bar H_{[6]}^\mu \, H_{[6]}^\nu\Big\} \,i\, \sigma_{\mu\nu}\,\mathcal{\bar H}  \,.
\label{LHQ}
\end{eqnarray}
We recall that the fields $\mathcal{H}$ and $H_{[6,\bar3]}$ are three-dimensional row and square matrices in flavor space, each of its components consisting of a 4 dimensional Dirac matrix. In addition, ${\rm Tr}$ and ${\rm tr}$\, stand for traces in Dirac and flavor spaces, respectively.

By using Eq.~(\ref{non-relativistic-expansion}) and Eq.~(\ref{phase-convention}), one can rewrite the chiral lagrangian in Eq.~(\ref{counter-terms}) and the $SU_v(2)$-invariant effective Lagrangian in Eq.~(\ref{LHQ}) in terms of the fields $P_+$\,, $P^\mu_+$\,, $B_{+}$\,, $B_{+}^{\mu}$\, and $B_{[\bar3]}^{+}$\,. Matching the structures, up to the leading order of $1/M_c$, between the non-relativistic expansion of the chiral Lagrangian and the heavy-quark spin symmetric Lagrangian, we obtain 35 sum rules
\begin{eqnarray}\label{largeMc-SumRule}
&& f_1^{(S)} = c_{1,[\bar3\bar3]}^{(S)} =  \tilde c_{1,[\bar3\bar3]}^{(S)}  \,,
\qquad
f_2^{(S)} = c_{2,[\bar3\bar3]}^{(S)} =  \tilde c_{2,[\bar3\bar3]}^{(S)} \,,
\nonumber\\
\nonumber\\
&& f_3^{(S)} = c_{1,[66]}^{(S)} =  \tilde c_{1,[66]}^{(S)} = d_{1,[66]}^{(S)} =  \tilde d_{1,[66]}^{(S)} \,,
\qquad
f_4^{(S)} = c_{2,[66]}^{(S)} =  \tilde c_{2,[66]}^{(S)} = d_{2,[66]}^{(S)} =  \tilde d_{2,[66]}^{(S)} \,,
\nonumber\\
\nonumber\\
&& f_1^{(A)} = e_{1,[\bar36]}^{(A)} = e_{2,[\bar36]}^{(A)} = \tilde e_{1,[\bar36]}^{(A)} = \tilde e_{2,[\bar36]}^{(A)}
= \sqrt3\,c_{1,[\bar36]}^{(A)} = \sqrt3\,\tilde c_{1,[\bar36]}^{(A)}  \,,
\nonumber\\
\nonumber\\
&&f_1^{(T)} = \tilde d_{1,[66]}^{(E)} = \tilde d_{3,[66]}^{(E)} = d_{1,[66]}^{(E)} = 3\, c_{1,[66]}^{(A)} = 3\,\tilde c_{1,[66]}^{(A)}
= \sqrt3\, e_{1,[66]}^{(A)} = \sqrt3\, e_{2,[66]}^{(A)} = \sqrt3\,\tilde e_{1,[66]}^{(A)} = \sqrt3\,\tilde e_{2,[66]}^{(A)}\,,
\nonumber\\
\nonumber\\
&& f_2^{(T)} = \tilde d_{2,[66]}^{(E)} = \tilde d_{4,[66]}^{(E)} = d_{2,[66]}^{(E)} = 3\, c_{2,[66]}^{(A)} = 3\,\tilde c_{2,[66]}^{(A)}
= \sqrt3\, e_{3,[66]}^{(A)} = \sqrt3\, e_{4,[66]}^{(A)} = \sqrt3\,\tilde e_{3,[66]}^{(A)} = \sqrt3\,\tilde e_{4,[66]}^{(A)} \,,
\nonumber\\
\nonumber\\
&& c_{1,[\bar36]}^{(S)} = \tilde c_{1,[\bar36]}^{(S)} = c_{1,[\bar3\bar3]}^{(A)} = c_{2,[\bar3\bar3]}^{(A)}  = \tilde c_{1,[\bar3\bar3]}^{(A)} = \tilde c_{2,[\bar3\bar3]}^{(A)}= 0\,.
\end{eqnarray}
According to the heavy-quark symmetry, the 42 parameters in the chiral Lagrangian with D-mesons and charmed baryons in Eq.~(\ref{counter-terms}) are reduced down to 7 independent parameters only.

\section{Large-$N_c$ operator analysis }
In this section, we employ the large-$N_c$ operator analysis to correlate the coupling constants. Here we use the main results and basic ideas from Refs. \cite{Luty:1993fu,Dashen:1993jt,Dashen:1994qi,Lutz:2010se,Lutz:2011fe,Lutz:2014jja}, where a formalism for the systematic expansion of baryon-matrix elements of QCD quark currents in powers of $1/N_c$\, was developed. Our strategy of this section is to first calculate the charmed baryon matrix elements of productions of two axial-vector or vector quark currents. The second step is to construct the operator product expansions of large-$N_c$ QCD effective operators bearing the structures of charmed baryon matrix elements. Finally, we match the structures of the two results to derive large-$N_c$ sum rules on the low-energy constants of the counter terms interactions in Eq.~(\ref{counter-terms}).

Our starting point is the correlation function of two axial-vector or vector quark currents, reading as \begin{eqnarray}
\bar C^{XY}_{\mu \nu,\,\bar a} (\bar q, q) =
\frac{\bar q^2 - M^2_X}{f_X}\,C^{XY}_{\mu \nu,\, \bar a} (\bar q, q)\, \frac{q^2 - M^2_Y}{f_Y}\,,
\label{def-Cbar}
\end{eqnarray}
with $X,Y = V,A$ and $M_A $ and $M_V$ the masses of the pseudo-scalar and vector D mesons in the flavor SU(3) limit.
In Eq. (\ref{def-Cbar}), we identify $q_\mu$ and $\bar q_\mu$ with the 4-momenta of the incoming and outgoing D mesons.
The $C^{XY}_{\mu \nu,\, \bar a} (\bar q, q)$ functions are given by :
\begin{eqnarray}
&& C^{AA}_{\mu \nu,a} (q) = i\,\int d^4 x \,e^{-i\,q\cdot x}  \,{\mathcal T}\, A_\mu (0)\,\lambda_{\bar a}\,\bar  A_\nu(x)\,,
\qquad \bar  A_\mu(x) = A^\dagger_\mu(x)\,,
\nonumber\\
&& C^{VV}_{\mu \nu,a} (q) = i\,\int d^4 x \,e^{-i\,q\cdot x}  \,{\mathcal T}\, V_\mu (0)\,\lambda_{\bar a}\,\bar V_\nu(x)\,,
\qquad \,\bar V_\mu(x) = V^\dagger_\mu(x)\,,
\nonumber\\
&& C^{VA}_{\mu \nu,a} (q) = i\,\int d^4 x \,e^{-i\,q\cdot x}  \,{\mathcal T}\, V_\mu (0)\,\lambda_{\bar a}\,\bar  A_\nu(x)\,,
\label{def-Cij}
\end{eqnarray}
with the quark field operators $u(x),d(x),s(x), c(x)$ of the up, down, strange and charm quarks where $\lambda_{\bar a}$ with $\bar a = 0, \cdots , 8$\, and ${\mathcal T}$\, is time ordering operator.
With the chiral Lagrangian in Eq.~(\ref{counter-terms}), we calculate the matrix elements of the correction functions in charmed baryon states at leading order of non-relativistic expansion by following the notations and conventions from \cite{Lutz:2011fe,Lutz:2014jja}. The physical charmed baryon states are defined by \cite{Lutz:2014jja},
\begin{eqnarray}
\ket{p,\, ij_\pm ,\,S,\,\chi }\,,
\label{def-states}
\end{eqnarray}
specified by the momentum $p$ and the flavor indices $i,j=1,2,3$, the spin $S$ and the spin-polarization
$\chi = 1,2$ for the spin one-half ($S=1/2$) and $\chi =1,\cdots ,4$ for the spin three-half states $(S=3/2)$. The
flavor sextet and the anti-triplet are discriminated by their symmetric (index $+$) and anti-symmetric (index $-$)
behaviour under the exchange of $i \leftrightarrow j$. More technical details of the calculation of functions $\bar C^{XY}_{\mu \nu,\,\bar a} (\bar q, q)$\, have been shown explicitly in Refs. \cite{Krause:1990xc,Lutz:2010se,Lutz:2011fe}. The leading terms in the low-momentum expansion of the matrix elements of the product of the two
axial-vector currents are
\begin{align}
\bra{\bar p, mn_+, \textstyle{\frac12}, \bar \chi}\, \bar C^{AA}_{ij,\bar a}\,  \ket{p, kl_+, \textstyle{\frac12}, \chi} & =  \bar p_i \, p_j\, \delta_{\bar \chi \chi}  \times \left\{
\begin{array}{l}
  \left( \sqrt{\frac 23} \,c^{(S)}_{1,[66]} + \sqrt{\frac 32}\,c^{(S)}_{2,[66]} \right) \delta_{(kl)_+}^{(mn)_+}
  \\
  c^{(S)}_{1,[66]} \,\Lambda_{(kl)_+}^{a,(mn)_+}  ,
\end{array}
\right. \nonumber \\ \nonumber \\
\bra{\bar p, mn_-, \textstyle{\frac12}, \bar \chi}\, \bar C^{AA}_{ij,\bar a}\, \ket{p, kl_-, \textstyle{\frac12}, \chi} &= \bar p_i \, p_j\, \delta_{\bar \chi \chi}  \times \left\{
\begin{array}{l}
  \left( \sqrt{\frac 23} \,c^{(S)}_{1,[\bar3\bar3]} + \sqrt{\frac 32}\,c^{(S)}_{2,[\bar3\bar3]} \right) \delta_{(kl)_-}^{(mn)_-}  \\
  c^{(S)}_{1,[\bar3\bar3]} \,\Lambda_{(kl)_-}^{a,(mn)_-}  ,
\end{array}
\right. \nonumber \\ \nonumber \\
\bra{\bar p, mn_+, \textstyle{\frac12}, \bar \chi}\, \bar C^{AA}_{ij,\bar a} \,\ket{p, kl_-, \textstyle{\frac12}, \chi} &=
-\,\bar p_i \,p_j\, \delta_{\bar \chi \chi}    \times \left\{
\begin{array}{l}
  0 \\
  c^{(S)}_{1,[\bar36]} \,\Lambda_{(kl)_-}^{a,(mn)_+} ,
\end{array}
\right.
\nonumber\\ \nonumber \\
\bra{\bar p, mn_+, \textstyle{\frac32}, \bar \chi}\, \bar C^{AA}_{ij,\bar a}\,  \ket{p, kl_+, \textstyle{\frac12}, \chi} & =  0, \nonumber \\
\nonumber \\
\bra{\bar p, mn_+, \textstyle{\frac32}, \bar \chi}\, \bar C^{AA}_{ij,\bar a} \,\ket{p, kl_-, \textstyle{\frac12}, \chi} &= 0, \nonumber \\
\nonumber \\
\bra{\bar p, mn_+, \textstyle{\frac32}, \bar \chi}\, \bar C^{AA}_{ij,\bar a} \,\ket{p, kl_+, \textstyle{\frac32}, \chi} &=  \bar p_i \,p_j\, \delta_{\bar \chi \chi}    \times \left\{
\begin{array}{l}
  \left(\sqrt{\frac 23} \,d^{(S)}_{1,[66]} + \sqrt{\frac 32} \, d^{(S)}_{2,[66]} \right) \delta_{(kl)_+}^{(mn)_+} \\
  d^{(S)}_{1,[66]} \,\Lambda_{(kl)_+}^{a,(mn)_+} .
\end{array}
\right.
\label{result:matrix_elements_low_energy_expansion_AA}
\end{align}
Here and in the following the upper row corresponds to the singlet component of the correlation function, $\bar C_{ij, 0}$, whereas the second row specifies the matrix elements of its octet components with $a=1,\ldots 8$. Furthermore, the flavor summation indices are $k,l,m,n=1,2,3$. Note that we have introduced in the calculation the convenient flavor structures \cite{Lutz:2014jja} :
\begin{eqnarray}
\delta_{kl_\pm}^{mn_\pm}
&=&  \frac 12\,\Big( \delta_{mk}\,\delta_{nl} \pm \delta_{nk}\,\delta_{ml} \,\Big)\,,
\nonumber\\
\Lambda_{kl_\pm}^{(a),\,mn_\pm}
&=&  \frac 14\,\Big( \lambda_{mk}^{(a)}\,\delta_{nl} \pm \lambda_{nk}^{(a)}\,\delta_{ml}
\pm \lambda_{ml}^{(a)}\,\delta_{nk} + \lambda_{nk}^{(a)}\,\delta_{ml} \,\Big)\,,
\nonumber\\
\Lambda_{kl_\pm}^{(a),\,mn_\mp}
&=&  \frac 14\,\Big( \lambda_{mk}^{(a)}\,\delta_{nl} \mp \lambda_{nk}^{(a)}\,\delta_{ml}
\pm \lambda_{ml}^{(a)}\,\delta_{nk} - \lambda_{nl}^{(a)}\,\delta_{mk} \,\Big)\,,
\label{def:flavor}
\end{eqnarray}
and spin properties
\begin{eqnarray}
&& S^\dagger_i\, S_j= \delta_{ij} - \frac{1}{3}\sigma_i \sigma_j \,, \qquad
S_i\,\sigma_j - S_j\,\sigma_i = -i\,\varepsilon_{ijk} \,S_k\,,
\qquad \vec S\cdot   \vec S^\dagger= \one_{(4\times 4)}\,,
\nonumber\\
&& \vec S^\dagger \cdot  \vec S =2\, \one_{(2\times 2)}\,, \qquad \vec S \cdot \vec \sigma = 0 \,,\qquad
\epsilon_{ijk}\,S_i\,S^\dagger_j = i\,\vec S \,\sigma_k\,\vec S^\dagger\,.
\label{def:spin}
\end{eqnarray}

The leading non-relativistic expansion of charmed baryon matrix elements for the product of two vector currents read
\allowdisplaybreaks
\begin{align}
\bra{\bar p, mn_+, \textstyle{\frac12}, \bar \chi}\, \bar C^{VV}_{ij,\bar a}\,\ket{p, kl_+, \textstyle{\frac12}, \chi}
 & =
\delta_{ij}\, \delta_{\bar \chi \chi}  \times \left\{
\begin{array}{l}
  \left( \sqrt{\frac 23} \,\tilde c^{(S)}_{1,[66]} + \sqrt{\frac 32} \,\tilde c^{(S)}_{2,[66]} \right) \delta_{(kl)_+}^{(mn)_+}\\
  \tilde c^{(S)}_{1,[66]}\,\Lambda_{(kl)_+}^{a,(mn)_+}
\end{array}
\right. \nonumber \\
& + i\epsilon_{ijk}\, \sigma^{k}_{\bar \chi \chi}  \times \left\{
\begin{array}{l}
  \left( \sqrt{\frac 23}\, \tilde c^{(A)}_{1,[66]} + \sqrt{\frac 32}\, \tilde c^{(A)}_{2,[66]} \right) \delta_{(kl)_+}^{(mn)_+}\\
  \tilde c^{(A)}_{1,[66]}\,\Lambda_{(kl)_+}^{a,(mn)_+},
\end{array}
\right. \nonumber \\ \nonumber \\
\bra{\bar p, mn_-, \textstyle{\frac12}, \bar \chi}\, \bar C^{VV}_{ij,\bar a} \,\ket{p, kl_-, \textstyle{\frac12}, \chi} &=  \delta_{ij}\, \delta_{\bar \chi \chi}  \times \left\{
\begin{array}{l}
  \left( \sqrt{\frac 23} \,\tilde c^{(S)}_{1,[\bar3\bar3]} + \sqrt{\frac 32} \,\tilde c^{(S)}_{2,[\bar3\bar3]} \right) \delta_{(kl)_-}^{(mn)_-}\\
  \tilde c^{(S)}_{1,[\bar3\bar3]}\,\Lambda_{(kl)_-}^{a,(mn)_-}
\end{array}
\right. \nonumber \\
& + i\epsilon_{ijk}\, \sigma^k_{\bar \chi \chi}  \times \left\{
\begin{array}{l}
  \left( \sqrt{\frac 23}\, \tilde c^{(A)}_{1,[\bar3\bar3]} + \sqrt{\frac 32}\, \tilde c^{(A)}_{2,[\bar3\bar3]} \right) \delta_{(kl)_-}^{(mn)_-}\\
  \tilde c^{(A)}_{1,[\bar3\bar3]}\,\Lambda_{(kl)_-}^{a,(mn)_-},
\end{array}
\right.  \nonumber \\ \nonumber \\
\bra{\bar p, mn_+, \textstyle{\frac12}, \bar \chi}\, \bar C^{VV}_{ij,\bar a} \,\ket{p, kl_-, \textstyle{\frac12}, \chi} &=
-\,\delta_{ij}\, \delta_{\bar \chi \chi}  \times \left\{
\begin{array}{l}
  0\\
  \tilde c^{(S)}_{1,[\bar36]}\,\Lambda_{(kl)_-}^{a,(mn)_-}
\end{array}
\right. \nonumber \\
&+\, \,i\epsilon_{ijk}\, \sigma^k_{\bar \chi \chi}  \times \left\{
\begin{array}{l}
  0\\
  \tilde c^{(A)}_{1,[\bar36]} \,\Lambda_{(kl)_-}^{a,(mn)_+},
\end{array}
\right.  \nonumber \\ \nonumber \\
\bra{\bar p, mn_+, \textstyle{\frac32}, \bar \chi}\, \bar C^{VV}_{ij,\bar a} \,\ket{p, kl_+, \textstyle{\frac12}, \chi} & =
\frac12\,\big(S_i \sigma_j \big)_{\bar \chi \chi}  \times \left\{
\begin{array}{l}
  \left( \sqrt{\frac 23}\, \tilde e^{(A)}_{1,[66]} + \sqrt{\frac 32}\, \tilde e^{(A)}_{3,[66]} \right) \delta_{(kl)_+}^{(mn)_+}\\
  \tilde e^{(A)}_{1,[66]}\,\Lambda_{(kl)_+}^{a,(mn)_+}
\end{array}
\right. \nonumber \\
& - \frac12\,\big(S_j \sigma_i \big)_{\bar \chi \chi}  \times \left\{
\begin{array}{l}
  \left( \sqrt{\frac 23}\, \tilde e^{(A)}_{2,[66]} + \sqrt{\frac 32}\, \tilde e^{(A)}_{4,[66]} \right) \delta_{(kl)_+}^{(mn)_+}\\
  \tilde e^{(A)}_{2,[66]}\,\Lambda_{(kl)_+}^{a,(mn)_+},
\end{array}
\right. \nonumber \\ \nonumber \\
\bra{\bar p, mn_+, \textstyle{\frac32}, \bar \chi}\, \bar C^{VV}_{ij,\bar a} \,\ket{p, kl_-, \textstyle{\frac12}, \chi} &=
 -\,\frac{1}{2}\,\big(S_i \sigma_j \big)_{\bar \chi \chi} \times
 \left\{
\begin{array}{l}
  0 \\
  \tilde e^{(A)}_{1,[\bar36]}\,\Lambda_{(kl)_-}^{a,(mn)_+}
\end{array}
\right. \nonumber \\
& +\, \frac{1}{2}\,\big(S_j\, \sigma_i \big)_{\bar \chi \chi} \times
 \left\{
\begin{array}{l}
  0 \\
  \tilde e^{(A)}_{2,[\bar36]}\,\Lambda_{(kl)_-}^{a,(mn)_+},
\end{array}
\right. \nonumber \\ \nonumber\\
\bra{\bar p, mn_+, \textstyle{\frac32}, \bar \chi}\, \bar C^{VV}_{ij,\bar a} \,\ket{p, kl_+, \textstyle{\frac32}, \chi} &=
\delta_{ij}\, \delta_{\bar \chi \chi} \times \left\{
\begin{array}{l}
  \left( \sqrt{\frac 23} \,\tilde d^{(S)}_{1,[66]} + \sqrt{\frac 32} \,\tilde d^{(S)}_{2,[66]} \right) \delta_{(kl)_+}^{(mn)_+}\\
  \tilde d^{(S)}_{1,[66]}\,\Lambda_{(kl)_+}^{a,(mn)_+}
\end{array}
\right. \nonumber \\
&  + \frac 12 \,\big(S_j S^\dagger_i \big)_{\bar \chi \chi}  \times \left\{
\begin{array}{l}
  \left(\sqrt{\frac 23} \,\tilde d^{(E)}_{1,[66]} + \sqrt{\frac 32} \, \tilde d^{(E)}_{2,[66]} \right) \delta_{(kl)_+}^{(mn)_+}\\
  \tilde d^{(E)}_{1,[66]} \,\Lambda_{(kl)_+}^{a,(mn)_+}
\end{array}
\right. \nonumber \\
& - \frac 12\, \big(S_i \,S^\dagger_j \big)_{\bar \chi \chi} \times \left\{
\begin{array}{l}
  \left(\sqrt{\frac 23} \,\tilde d^{(E)}_{3,[66]} + \sqrt{\frac 32} \, \tilde d^{(E)}_{4,[66]} \right) \delta_{(kl)_+}^{(mn)_+}\\
  \tilde d^{(E)}_{3,[66]} \,\Lambda_{(kl)_+}^{a,(mn)_+}.
\end{array}
\right.
\label{result:matrix_elements_low_energy_expansion_VV}
\end{align}
The leading non-relativistic expansion of charmed baryon matrix elements for the product of a vector and an axial-vector currents is derived as
\begin{align}
\bra{\bar p, mn_+, \textstyle{\frac12}, \bar \chi}\, \bar C^{VA}_{ij,\bar a}\,\ket{p, kl_+, \textstyle{\frac12}, \chi}  & =
  -\,p_j\, (\sigma_i)_{\bar \chi \chi}  \times \left\{
\begin{array}{l}
  \left( \sqrt{\frac 23} c^{(A)}_{1,[66]} + \sqrt{\frac 32} \,c^{(A)}_{2,[66]} \right) \delta_{(kl)_+}^{(mn)_+}\\
  c^{(A)}_{1,[66]}\,\Lambda_{(kl)_+}^{a,(mn)_+},
\end{array}
\right. \nonumber \\ \nonumber\\
\bra{\bar p, mn_-, \textstyle{\frac12}, \bar \chi}\, \bar C^{VA}_{ij,\bar a} \,\ket{p, kl_-, \textstyle{\frac12}, \chi} &=
-\,p_j\, (\sigma_i)_{\bar \chi \chi}  \times \left\{
\begin{array}{l}
  \left( \sqrt{\frac 23} c^{(A)}_{1,[\bar3\bar3]} + \sqrt{\frac 32} \,c^{(A)}_{2,[\bar3\bar3]} \right) \delta_{(kl)_-}^{(mn)_-}\\
  c^{(A)}_{1,[\bar3\bar3]}\,\Lambda_{(kl)_-}^{a,(mn)_-},
\end{array}
\right. \nonumber \\ \nonumber\\
\bra{\bar p, mn_+, \textstyle{\frac12}, \bar \chi}\, \bar C^{VA}_{ij,\bar a} \,\ket{p, kl_-, \textstyle{\frac12}, \chi} &=
-\, p_j\, (\sigma_i)_{\bar \chi \chi}  \times \left\{
\begin{array}{l}
  0\\
  c^{(A)}_{1,[\bar36]}\,\Lambda_{(kl)_-}^{a,(mn)_+},
\end{array}
\right.
\nonumber\\ \nonumber\\
\bra{\bar p, mn_+, \textstyle{\frac32}, \bar \chi}\, \bar C^{VA}_{ij,\bar a} \,\ket{p, kl_+, \textstyle{\frac12}, \chi} & =
  \frac12\,p_j\, (S_i)_{\bar \chi \chi}  \times \left\{
\begin{array}{l}
  \left( \sqrt{\frac 23} e^{(A)}_{1,[66]} + \sqrt{\frac 32} \,e^{(A)}_{3,[66]} \right) \delta_{(kl)_+}^{(mn)_+}\\
  e^{(A)}_{1,[66]}\,\Lambda_{(kl)_+}^{a,(mn)_+},
\end{array}
\right. \nonumber \\ \nonumber\\
\bra{\bar p, mn_+, \textstyle{\frac32}, \bar \chi}\, \bar C^{VA}_{ij,\bar a} \,\ket{p, kl_-, \textstyle{\frac12}, \chi} &=
 -\,\frac{1}{2}\, p_j\, (S_i)_{\bar \chi \chi}  \times \left\{
\begin{array}{l}
  0\\
  e^{(A)}_{1,[\bar36]}\,\Lambda_{(kl)_-}^{a,(mn)_+},
\end{array}
\right. \nonumber\\ \nonumber\\
\bra{\bar p, mn_+, \textstyle{\frac32}, \bar \chi}\, \bar C^{VA}_{ij,\bar a} \,\ket{p, kl_+, \textstyle{\frac32}, \chi} &=
-\frac 12\, p_j\, \big(\vec S \sigma_i\, \vec S^\dagger \big)_{\bar \chi \chi} \times \left\{
\begin{array}{l}
  \left( \sqrt{\frac 23} d^{(E)}_{1,[66]} + \sqrt{\frac 32} \,d^{(E)}_{2,[66]} \right) \delta_{(kl)_+}^{(mn)_+}\\
  d^{(E)}_{1,[66]}\,\Lambda_{(kl)_+}^{a,(mn)_+}\,.
\end{array}
\right.
\label{result:matrix_elements_low_energy_expansion_VA}
\end{align}
The charmed baryon matrix elements of the correlation functions in Eqs.~(\ref{result:matrix_elements_low_energy_expansion_AA}, \ref{result:matrix_elements_low_energy_expansion_VV}, \ref{result:matrix_elements_low_energy_expansion_VA}) will be used to match the spin and flavor structures with the large-$N_c$ effective operator product expansion which will be worked out in the following.

According to Refs. \cite{Luty:1993fu,Dashen:1994qi}, the spin-flavor symmetry of large-$N_c$ baryon analysis allows us to perform a systematic $1/N_c$ expansion for baryon matrix elements. The $1/N_c$ expansion of the correlation functions in c.m. frame takes a generic form \cite{Lutz:2010se,Lutz:2011fe}
\begin{eqnarray}
\bra{\bar p,\,\bar \chi} \, \bar C^{}_{\mu \nu,a}(\bar q,q)\, \ket{p,\,\chi} = \sum_r
c_r (\bar p, p) \roundbra{\bar \chi} \, \mathcal{O}^{(r)}_{\mu \nu,a} \,\roundket{\chi}\,.
\label{def-expansion}
\end{eqnarray}
All dynamical information of the correlation functions with physical states is transferred into the unknown coefficient functions, $c_r (\bar p, p)$\,, the $\roundket{\chi}$ states reflect spin and flavor structures only.
The effective operators $\mathcal{O}^{(r)}_{\mu \nu,a}$ can be written in terms of $\,(J)^l\,(T)^m\,(G)^n$\, with $\,l+m+n = r$\,. The operators $J$\,, $T$\, and $G$\, are spin, flavor and spin-flavor operators, respectively, and the detail definitions are referred to \cite{Dashen:1994qi,Lutz:2010se}.
The unknown coefficient functions have the $N_c$ scaling as
\begin{eqnarray}
c_r \sim \frac{1}{N_c^{r-1}}\,,
\label{cr-order}
\end{eqnarray}
since the matrix elements of correlation functions in physical states have the scale $\mathcal{O}(N_c)$ but the matrix elements of $\mathcal{O}^{(r)}$ operators scale as $N_c^r$\,.

The effective operators have the $N_c$ scaling as follows \cite{Dashen:1993jt}:
\begin{eqnarray}
J^i \sim \frac{1}{N_c} \,, \qquad \quad T^{a} \sim N^0_c \,, \qquad \quad G^{i\,a} \sim N^0_c \,.
\label{effective-counting}
\end{eqnarray}
The $N_c$ scaling rules in Eq.~(\ref{effective-counting}) alone can not eliminate redundant operators at a given order of $1/N_c$. 
However, the operator identities with the scaling rules together allow a systematic summation of the relevant operators at a given $1/N_c$ order. The operator identities are first derived in \cite{Dashen:1993jt} for the light-quark operators, and later generalized to baryons containing the heavy-quarks \cite{Jenkins:1996de}. The structure of operator identities and its derivation for the baryon containing the light- and heavy-quarks are given in technical detail in Refs. \cite{Jenkins:1996de,Lutz:2014jja}. We use only the main results here.

In this work, our effective charmed baryon states, $\roundket{ij_\pm\,,\,\chi}$, are composed of light- and heavy-quarks. It has been demonstrated by Ref. \cite{Jenkins:1996de} that one can generalize the $1/N_c$ expansion in Eq.~(\ref{def-expansion}) from light quark baryons to light and heavy quark systems. The effective $r$-operator can be written in the following form
\begin{eqnarray}
\mathcal{O}^{(r)} = \mathcal{O}_{\rm light}^{(p)}\,\mathcal{O}_{\rm heavy}^{(q)}\,,
\end{eqnarray}
with $r = p+q$\,, where $\mathcal{O}_{\rm light}^{(p)}$\, and $\mathcal{O}_{\rm heavy}^{(q)}\,$ are the effective light quark $p$-body and heavy quark $q$-body operators, respectively. In this study, we consider baryons containing a single charm quark and keep the heavy-quark spin symmetry. The effective operators for $\mathcal{O}_{\rm heavy}^{(q)}$ could be only the heavy-quark spin operator, $J_Q^i$
since the heavy-quark flavor and spin-flavor operators are irrelevant.
However, the operator $J_Q^i \sim 1/M_Q$ breaks the heavy-quark spin symmetry \cite{Jenkins:1996de,Lutz:2014jja}, and therefore $\mathcal{O}_{\rm heavy}^{(q)}$ should be trivial in our consideration.

By applying the operator identities in Refs. \cite{Jenkins:1996de,Lutz:2014jja} with the operators scaling in Eq.~(\ref{effective-counting}) and the ansatz for the operator production expansion in Ref. \cite{Lutz:2011fe}, we find 4 operators at leading order $\mathcal{O}\big( N_c^{0}\big)$
\begin{eqnarray}
\bar p_i\, p_j\, T^{\bar a}\,,\qquad \delta_{ij}\,T^{\bar a}\,,\qquad \epsilon_{ijk}\, G^{\bar a,k}\,,\qquad
p_j\, G^{\bar a}_i\,,
\end{eqnarray}
and  7 operators at $\mathcal{O}\big( N_c^{-1}\big)$
\begin{eqnarray}
&& \bar p_i\, p_j\,\big[J^k, \,G_k^{\bar a}\big]_+\,,\qquad \delta_{ij}\,\big[J^k, \,G_k^{\bar a}\big]_+\,,\qquad
\epsilon_{ijk}\,\big[J^k, \,T^{\bar a}\big]_+\,,\qquad \big[J_i, \,G_j^{\bar a}\big]_+\,,\quad \big[J_j, \,G_i^{\bar a}\big]_+\,,
\nonumber\\
&& p_j\,\big[J_i, \,T^{\bar a}\big]_+\,,\qquad \varepsilon_{ikl}\, \big[J_k, \,G^{\bar a}_l\big]_+ \,.
\end{eqnarray}
 At the next-leading order in the power of $1/N_c$ expansion, the correlation functions in Eq.~(\ref{def-Cbar}) can be written in terms of the 11 operators above,
\begin{align}
\bra{\bar p, \bar \chi}  \,\bar C^{AA}_{ij,\bar a}\, \ket{p, \chi}  & = \bar p_i\, p_j\,
\roundbra{\bar \chi} \, g^{AA}_1\, T^{\bar a}
 + \smallfrac 12 \,g^{AA}_2\, \big[J^k, \,G_k^{\bar a}\big]_+  \roundket{\chi},
\nonumber\\
\langle \bar p, \bar \chi |  \,\bar C^{VV}_{ij,{\bar a}}\,| p, \chi  \rangle  &=
 \delta_{ij}\, ( \bar \chi | \, g^{VV}_1\,T^{\bar a} + \smallfrac{1}{2}\,g^{VV}_2\, \big[J^k, \,G_k^{\bar a}\big]_+
 \roundket{\chi} \nonumber \\
&+ i\, \epsilon_{ijk} \roundbra{\bar \chi }\, g^{VV}_3\, G^{\bar a,k} + \smallfrac{1}{2}\,g^{VV}_4\, \big[J^k, \,T^{\bar a}\big]_+
\roundket{\chi} \nonumber \\
&+ \roundbra{\bar \chi} \smallfrac{1}{2}\,g^{VV}_5\, \big[J_i, \,G_j^{\bar a}\big]_+ +
\smallfrac{1}{2}\,g^{VV}_6\, \big[J_j, \,G_i^{\bar a}\big]_+ \roundket{\chi},
\nonumber\\
\langle \bar p, \bar \chi |  \,\bar C^{VA}_{ij,{\bar a}}\,| p, \chi  \rangle  & =
p_j\, ( \bar \chi | \, g^{VA}_1\,  G^{\bar a}_i + \smallfrac{1}{2}\,g^{VA}_2\, \big[J_i, \,T^{\bar a}\big]_+ +
\smallfrac{1}{2}\,g^{VA}_3\, i\, \varepsilon_{ikl}\, \big[J_k, \,G^{\bar a}_l\big]_+ \roundket{\chi} .
\label{result:matrix_elements_Nc_expansion}
\end{align}
Note that in the calculation above we have used the results of the actions of one-body and all symmetric combinations of two one-body effective operators on effective charmed baryons states with truncation at $N_c=3$ in Ref. \cite{Lutz:2014jja}. It is found that the coupling constants of effective operators in Eq.~(\ref{result:matrix_elements_Nc_expansion}) have the $N_c$ scaling as follows :
\begin{eqnarray}
g_1^{AA}\,,\, g_{1,3}^{VV}\,,\, g_1^{VA}\, \sim\, N_c^0\,,\qquad
g_2^{AA}\,,\, g_{2,4,5,6}^{VV}\,,\, g_{2,3}^{VA}\, \sim\, N_c^{-1}\,.
\label{g-eff-Nc}
\end{eqnarray}

Matching the spin-flavor structures between the non-relativistic expansions in Eqs.~(\ref{result:matrix_elements_low_energy_expansion_AA}), (\ref{result:matrix_elements_low_energy_expansion_VV}) and
(\ref{result:matrix_elements_low_energy_expansion_VA}) and the charmed baryon matrix elements of the $1/N_c$ effective quark operator product expansions in Eq.~(\ref{result:matrix_elements_Nc_expansion}) up to order $\mathcal{O}\big( N_c^{-1}\big)$ leads to the following correlations between the parameters in both expansions,
\allowdisplaybreaks
\begin{eqnarray}
c_{1,[66]}^{(S)} &=& g_1^{AA} + g_2^{AA}\,,\quad c_{2,[66]}^{(S)} = 0\,,\quad c_{1,[\bar3\bar3]}^{(S)} = g_1^{AA}\,,\quad c_{2,[\bar3\bar3]}^{(S)} = 0\,, \quad c_{1,[\bar36]}^{(S)} = 0\,,
\nonumber\\
d_{1,[66]}^{(S)} &=& g_1^{AA} + g_2^{AA}\,,\quad d_{2,[66]}^{(S)} = 0\,,
\nonumber\\
\nonumber\\
\tilde c_{1,[66]}^{(S)} &=& g_1^{VV} + g_2^{VV}\,,\quad c_{2,[66]}^{(S)} = 0\,,\quad \tilde c_{1,[66]}^{(A)} = -\frac 23\,g_3^{VV} -\frac 43\,g_4^{VV}\,,\quad \tilde c_{2,[66]}^{(A)} = 0\,,
\nonumber\\
\tilde c_{1,[\bar3\bar3]}^{(S)} &=& g_1^{VV}\,,\quad
\tilde c_{2,[\bar3\bar3]}^{(S)} =0\,,\quad \tilde c_{1,[\bar3\bar3]}^{(A)} = 0\,,\quad
\tilde c_{2,[\bar3\bar3]}^{(A)} =0\,,\quad
\nonumber\\
\tilde c_{1,[\bar36]}^{(S)} &=& 0\,,\quad \tilde c_{1,[\bar36]}^{(A)} = -\frac{1}{\sqrt 3}\,g_3^{VV} +\frac{1}{2\sqrt{3}}\,g_-\,,
\nonumber\\
\tilde e_{1,[66]}^{A} &=& -\frac{1}{\sqrt 3}\left( g_3^{VV} + 2\,g_4^{VV} \right) - \frac{1}{2\sqrt{3}}\,g_+\,,\quad \tilde e_{2,[66]}^{A} = -\frac{1}{\sqrt 3}\left( g_3^{VV} + 2\,g_4^{VV} \right) + \frac{1}{2\sqrt{3}}\,g_+\,,
\nonumber\\
\tilde e_{3,[66]}^{A} &=& 0\,,\quad \tilde e_{4,[66]}^{A} = 0\,,\quad \tilde e_{1,[\bar36]}^{A} = -g_3^{VV} - \frac{1}{2}\,g_+\,,\quad \tilde e_{2,[\bar36]}^{A} = -g_3^{VV} - \frac{1}{2}\,g_+\,,
\nonumber\\
\tilde d_{1,[66]}^{(S)} &=& g_1^{VV} + g_2^{VV} + \frac{1}{2}\,g_+\,,\quad \tilde d_{2,[66]}^{(S)}  = 0\,,\quad \tilde d_{1,[66]}^{(E)} = -\left( g_3^{VV} + 2\,g_4^{VV} \right) - \frac{1}{2}\,g_+\,,\quad \tilde d_{2,[66]}^{(E)} = 0\,,
\nonumber\\
\tilde d_{3,[66]}^{(E)} &=& -\left( g_3^{VV} + 2\,g_4^{VV} \right) + \frac{1}{2}\,g_+\,,\quad \tilde d_{4,[66]}^{(E)}  = 0\,,
\nonumber\\
\nonumber\\
c_{1,[66]}^{(A)} &=& -\frac{1}{3}\left( g_3^{VV} + 2\,g_4^{VV} \right),\quad
c_{1,[\bar36]}^{(A)} = \frac{1}{\sqrt3}\,g_1^{VA} - \frac{1}{\sqrt3}\,g_3^{VA}\,,
\nonumber\\
c_{1,[\bar3\bar3]}^{(A)} &=& 0\,,\quad c_{2,[\bar3\bar3]}^{(A)} = 0\,,
\nonumber\\
e_{1,[66]}^{(A)} &=& e_{2,[66]}^{(A)} =  -\frac{1}{3}\left( g_1^{VA} + 2\,g_2^{VA} \right),\quad e_{3,[66]}^{(A)} = e_{4,[66]}^{(A)} = 0\,,\quad
e_{1,[\bar36]}^{(A)} = e_{2,[\bar36]}^{(A)} = g_1^{VA} - g_3^{VA}\,,
\nonumber\\
d_{1,[66]}^{(E)} &=& -g_1^{VA} - 2\,g_2^{VA}\,,\quad  d_{2,[66]}^{(E)} = 0\,,
\label{matching}
\end{eqnarray}
where $g_{\pm} = g_5^{VV} \pm g_6^{VV}$\,.

Comparing with the 42 chiral parameters in the Lagrangian, we obtain a set of 29 sum rules,
\begin{eqnarray}\label{largeNc-sumrules}
c_{1,[66]}^{(S)} &=& d_{1,[66]}^{(S)}\,,\quad c_{1,[\bar36]}^{(S)} = 0\,, \quad c_{2,[66]}^{(S)} = 0\,, \quad
c_{2,[\bar3\bar3]}^{(S)} = 0\,, \quad d_{2,[66]}^{(S)} = 0\,,
\nonumber\\
\nonumber\\
\tilde c_{1,[66]}^{(S)} &=& \tilde d_{1,[66]}^{(S)} + \frac{1}{2\sqrt3}\,\tilde e_{1,[66]}^{(A)} - \frac{1}{2\sqrt3}\,\tilde e_{2,[66]}^{(A)}\,,\quad \tilde c_{1,[66]}^{(A)} = \frac{1}{2\sqrt3}\,\tilde e_{1,[66]}^{(A)} + \frac{1}{2\sqrt3}\,\tilde e_{2,[66]}^{(A)}\,,\quad \nonumber\\
\tilde c_{1,[\bar36]}^{(A)} &=& \frac{1}{\sqrt3}\,\tilde e_{2,[\bar36]}^{(A)}\,,\quad \tilde d_{1,[66]}^{(E)} = \sqrt{3}\,\tilde e_{1,[66]}^{(A)}\,,\quad \tilde d_{3,[66]}^{(E)} = \sqrt{3}\,\tilde e_{2,[66]}^{(A)}\,,\quad
\tilde e_{1,[\bar36]}^{(A)} = \tilde e_{2,[\bar36]}^{(A)}\,,
\nonumber\\
\tilde c_{1,[\bar3\bar3]}^{(A)} &=& 0\,, \quad \tilde c_{2,[66]}^{(S)} = 0\,, \quad \tilde c_{2,[\bar3\bar3]}^{(S)} = 0\,, \quad \tilde c_{2,[66]}^{(A)} = 0\,, \quad \tilde c_{2,[\bar3\bar3]}^{(A)} = 0 \,, \quad
\nonumber\\
\tilde d_{2,[66]}^{(S)} &=& 0\,, \quad \tilde d_{2,[66]}^{(E)} = 0\,, \quad \tilde d_{4,[66]}^{(E)} = 0\,, \quad \tilde e_{3,[66]}^{(A)} = 0\,, \quad \tilde e_{4,[66]}^{(A)} = 0\,,
\nonumber\\
\nonumber\\
c_{1,[66]}^{(A)} &=& \frac{1}{\sqrt 3}\,e_{1,[66]}^{(A)} \,,\quad d_{1,[66]}^{(E)} = \sqrt{3}\,e_{1,[66]}^{(A)} \,, \quad c_{1,[\bar 36]}^{(A)} = \frac{1}{\sqrt{3}}\,e_{1,[\bar 36]}^{(A)} \,,
\nonumber\\
c_{1,[\bar3\bar3]}^{(A)} &=& 0\,, \quad c_{2,[66]}^{(A)} = 0\,, \quad c_{2,[\bar3\bar3]}^{(A)} = 0\,, \quad d_{2,[66]}^{(E)} = 0\,, \quad e_{4,[66]}^{(A)} = 0 \,.
\end{eqnarray}
Considering only the large-$N_c$ operator expansion analysis, the free parameters in the Lagrangian are reduced from 42 to 42 $-$ 29 = 13. Therefore we conclude that large-$N_c$ analysis with $N_c=3$ determines 29 sum rules up to $\approx\, 30\,\%$ corrections.

The combination of the large-$N_c$ sum rules Eq. (\ref{largeNc-sumrules}) with the ones from the heavy-quark symmetry in Eq.~(\ref{largeMc-SumRule}) leads to 3 additional sum rules,
\begin{eqnarray}
&& f_2^{(S)} =  f_4^{(S)} =  f_2^{(T)} = 0\,.
\end{eqnarray}
Therefore, the independent sum rules are 38 totally and hence the coupling constants of chiral Lagrangian are reduced to $42-38=4$ free parameters only. The Lagrangian in Eq.~(\ref{counter-terms}) takes the form in terms of the 4 independent parameters,
\begin{eqnarray}
&& {\mathcal L}^{\rm counter} = g_1\left(D\,\bar B_{[\bar3]}\,B_{[\bar3]}\,\bar D
 -\, \frac{1}{2}\,D_{\mu \nu}\,\bar B_{[\bar3]}\,B_{[\bar3]}\,\bar D^{\mu\nu}\right)
\nonumber\\
&& \qquad +\, g_2\Bigg( D\,\bar B_{[6]}\,B_{[6]}\,\bar D -\, \frac{1}{2}\, D_{\mu\nu}\,\bar B_{[6]}\,B_{[6]}\,\bar D^{\mu\nu}
- D\,\bar B_{[6]}^\alpha\,g_{\alpha\beta} \, B_{[6]}^\beta\,\bar D
+\, \frac{1}{2}\,D_{\mu \nu}\,\bar B_{[6]}^\alpha\,g_{\alpha\beta} \, B_{[6]}^\beta\,\bar D^{\mu\nu} \Bigg)
\nonumber \\
&& \qquad +\, \frac{g_3}{2}\Bigg(- \, \frac{i}{\sqrt{3}\,M_c}\,\Big\{
D_{\mu \nu}\,\bar B_{[6]}\, \gamma^\mu\gamma_5\,\,B_{[\bar3]}\,(\partial^\nu \bar D)
- (\partial^\nu D)\,\bar B_{[6]}\, \gamma^\mu\gamma_5\,\,B_{[\bar3]}\,\bar D_{\mu \nu}\, \Big\}
+ {\rm h.c.}
\nonumber\\
&& \qquad\qquad\quad - \,\frac{1}{2\sqrt{3} \,M_c}\, \epsilon^{\mu \nu \alpha \beta}
\Big\{D_{\mu \nu} \,\bar B_{[6]}\, \gamma_\alpha\,\gamma_5 \,B_{[\bar3]}
(\partial^\tau \bar D_{\tau \beta} )
+ (\partial^\tau D_{\tau \beta} )\,\bar B_{[6]} \, \gamma_\alpha\,\gamma_5 \,B_{[\bar3 ]}
\,\bar D_{\mu\nu} \Big\} + {\rm h.c.}
\nonumber\\
&&\qquad\qquad\quad + \,\frac{i}{2}\,\epsilon_{\mu \nu \alpha \beta}\Big\{
D^{\alpha \beta} \, \bar B_{[6]}^\mu \,\gamma^{\nu}\, \gamma_5\,B_{[\bar3]} \, \bar D
+ D\,\bar B_{[6]}^\mu \,\gamma^{\nu}\,\gamma_5\,B_{[\bar3]} \, \bar D^{\alpha \beta} \Big\} + {\rm h.c.}
\nonumber\\
&&\qquad\qquad\quad + \,\Big\{ D_{\alpha \nu} \,
\bar B_{[6]}^\mu \, \gamma^{\nu}\,\gamma_5\,B_{[\bar3]}\, g^{\alpha\beta}\,\bar D_{\beta\mu}
- D_{\alpha \mu} \, \bar B_{[6]}^\mu \, \gamma^{\nu}\,\gamma_5\,B_{[\bar3]} \,g^{\alpha\beta}\,\bar D_{\beta \nu} \Big\} + {\rm h.c.}  \Bigg)
\nonumber \\
&& \qquad + \, g_4\Bigg( \frac{i}{M_c}\,D_{\mu \nu}\,\bar B_{[6]}\, \gamma^\mu\,\gamma_5\,B_{[6]}\,(\partial^\nu \bar D)
+ \,\frac{1}{4 \,M_c}\, \epsilon^{\mu \nu \alpha \beta}
D_{\mu \nu} \,\bar B_{[6]}\, \gamma_\alpha\,\gamma_5 \,B_{[6]}(\partial^\tau \bar D_{\tau \beta} ) + {\rm h.c.}
\nonumber\\
&&\qquad\qquad\quad + \frac{i}{4\sqrt{3}}\,\epsilon_{\mu \nu \alpha \beta}\,\Big\{\,D^{\alpha \beta} \,
\bar B_{[6]}^\mu \,\gamma^{\nu}\, \gamma_5\,B_{[6]} \, \bar D
+ D\,\bar B_{[6]}^\mu \,\gamma^{\nu}\,\gamma_5\,B_{[6]} \, \bar D^{\alpha \beta} \Big\} + {\rm h.c.}
\nonumber\\
&&\qquad\qquad\quad + \frac{1}{2\sqrt{3}}\,\Big\{ D_{\alpha \nu} \,
\bar B_{[6]}^\mu \, \gamma^{\nu}\,\gamma_5\,B_{[6]}\, g^{\alpha\beta}\,\bar D_{\beta\mu}
- D_{\alpha \mu} \,
\bar B_{[6]}^\mu \, \gamma^{\nu}\,\gamma_5\,B_{[6]} \,g^{\alpha\beta}\,\bar D_{\beta \nu} \Big\}  + {\rm h.c.}
\nonumber\\
&& \qquad\qquad\quad + \, \frac{i}{12 }\,\epsilon_{\mu \nu \alpha \beta }\,D^{\mu \nu}\,\bar B_{[6]}^\alpha\, B_{[6]}^\beta\,\bar D
 + {\rm h.c.}
\nonumber\\
&& \qquad\qquad\quad + \, \frac{1}{6}\,\Big\{D_{\beta \mu}\,\bar B_{[6]}^{\tau}\, B_{[6]}^\beta
\,\bar D_\tau^{~\,\mu} - D_{\alpha \mu}\,\bar B_{[6]}^{\alpha}\, B_{[6]}^\tau
\,\bar D_\tau^{~\,\mu}\Big\}\Bigg)\,.
\label{final-L}
\end{eqnarray}
Notice that the four free parameters have the $N_c$ scaling, $g_{1}\sim N_c^0$ and $g_{2,3,4} \sim N_c^0 + 1/N_c$\,, as shown in Eq.~(\ref{matching}) which match the chiral parameters with the coefficients of the effective operators in Eq.~(\ref{g-eff-Nc}).

The final lagrangian in Eq. (\ref{final-L}) implies that the $D\bar B_{[\bar3]}B_{[\bar3]}\bar D$ and $D^*\bar B_{[\bar3]}B_{[\bar3]}\bar D^*$ systems have the same scalar coupling $g_1$, while the $D\bar B_{[6]}B_{[6]}\bar D$, $D^*\bar B_{[6]}B_{[6]}\bar D^*$, $D\bar B_{[6]}^*B_{[6]}^*\bar D$ and $D^*\bar B_{[6]}^*B_{[6]}^*\bar D^*$ systems have the same coupling $g_2$ in the approximation of the heavy-quark symmetry and large-$N_c$. For the axial-vector interaction, the systems $D\bar B_{[6]}B_{[\bar 3]}\bar D^*$, $D^*\bar B_{[6]}B_{[\bar3]}\bar D^*$, $D\bar B_{[6]}^*B_{[\bar 3]}\bar D^*$ and $D^*\bar B_{[6]}^*B_{[\bar3]}\bar D^*$ have the same coupling $g_3$, but the systems $D\bar B_{[6]}B_{[6]}\bar D^*$, $D^*\bar B_{[6]}B_{[6]}\bar D^*$, $D\bar B_{[6]}^*B_{[6]}\bar D^*$ and $D^*\bar B_{[6]}^*B_{[6]}\bar D^*$ have the same coupling $g_4$.

\section{Summary}

We have constructed chiral $SU(3)$ Lagrangian with
$D$ mesons of spin $J^P=0^-$ and $J^P=1^-$ and charmed baryons of spin $J^P=1/2^+$ and $J^P=3/2^+$. There are altogether 42
independent terms in the chiral Lagrangian, which contribute, at chiral order $Q^0$, to scattering processes of the $D$ mesons and charmed baryons. We have applied both the heavy-quark symmetry and the large-$N_c$ analysis to constraint the coupling constants. The heavy-quark mass expansion provides 35 sum rules at leading order, while the $1/N_c$ analysis at next-leading order of $1/N_c$ expansion predicts 29 sum rules. The heavy-quark symmetry and large-$N_c$ operator analysis together arise totally 38 sum rules, and hence the unknown parameters in the Lagrangian are reduced down to 42$-$38 = 4 only, reduced by a factor of 10 approximately. The result remarkably demonstrates the consistence of the heavy-quark symmetry and the large-$N_c$ analysis for the chiral Lagrangian with $D$ mesons and charmed baryons. The sum rules are useful constraints in establishing a systematic coupled-channel chiral effective field theory for the $D$ meson and charmed-baryon scattering beyond the threshold region in the hidden and doubly charmed baryon systems.

Recently, LHCb has reported two hidden charmed baryon resonances, $P_c^+(4380)$ and $P_c^+(4450)$\, \cite{Aaij:2015tga}. However, whether these states are pentaquarks or molecules is still in question. The couple channel study of $D$ meson and charmed baryon systems in the short range interactions in this work and long range interactions \cite{Hofmann-Lutz-2005,Hofmann-Lutz-2006,Lutz:2006ya,Wu:2010jy,Wu:2010vk,Xiao:2013yca} may provide more information of properties of these two states.

The counter terms may serve as a good approximation to the contributions of various $t$-channel $\psi$ meson exchanges in charm meson and charmed baryon collisions. The effective coupling constants may be written in the form, for example, for the $DD\Lambda_c\Lambda_c$ counter term,
\begin{eqnarray}
g_{DD\Lambda_c\Lambda_c} = \sum_i g_{DD\psi_i}\frac{1}{M_{\psi_i}}g_{\psi_i\Lambda_c\Lambda_c}\,.
\end{eqnarray}
The coupling constants $g_{DD\psi_i}$ may be estimated by fitting $e^-e^+\rightarrow D\bar D$ data in meson-exchange theories \cite{Zhang:2009gy,Chen:2011xk,Achasov:2012ss} or in quark models \cite{Limphirat:2013jga}. There is no experimental data available for extracting the coupling constant $g_{\psi_i\Lambda_c\Lambda_c}$, but one may estimate it indirectly in the $^3P_0$ quark model \cite{Yan:2009vr}, considering that the $^3P_0$ quark dynamics is of independence of environments
where heavy quarks may or may not be a component of baryons \cite{Limphirat:2010zz}.
\section*{Acknowledgement}

This work is supported by Thailand research fund TRF-RMUTI under contract No. TRG5680079. DS and YY acknowledge support from Suranaree University of Technology (SUT) and the Office of the Higher Education Commission under NRU project of Thailand (SUT-COE: High Energy Physics \& Astrophysics) and SUT-CHE-NRU (NV11/2558). We are graceful to Matthias F. M. Lutz for persuading us to investigate on this work and valuable comments. DS thanks Pakakaew Rittipruk for providing useful references.


\begin{thebibliography}{9}
%
%
%




\bibitem{Hofmann-Lutz-2004}
J. Hofmann and M.F.M. Lutz, Nucl. Phys. {\bf A 733} (2004) 142.

\bibitem{Lutz-Soyeur-2006}
M.F.M. Lutz and M. Soyeur, Nucl. Phys. {\bf A 713} (2008) 14.

\bibitem{Guo-Hanhart-Krewald-Meissner-2008}
F.-K. Guo, C. Hanhart, S. Krewald, U.-G. Mei\ss ner
Phys. Lett. {\bf B 666} (2008) 251.

\bibitem{Kolomeitsev-Lutz-2004}
E.E. Kolomeitsev and M.F.M. Lutz, Phys. Lett. {\bf B 582} (2004)
39.

\bibitem{Lutz-Kolomeitsev-2004-charm}
M.F.M. Lutz and E.E. Kolomeitsev, Nucl. Phys. {\bf A 730} (2004) 110.

\bibitem{Lutz:2005ip}
  M.~F.~M.~Lutz and E.~E.~Kolomeitsev,
  Nucl.\ Phys.\ A {\bf 755}, 29 (2005).

\bibitem{Hofmann-Lutz-2005}
J. Hofmann and M.F.M. Lutz, Nucl. Phys. {\bf A 763} (2005) 90.

\bibitem{Hofmann-Lutz-2006}
J. Hofmann and M.F.M. Lutz, Nucl. Phys. {\bf A 776} (2006) 17.

\bibitem{Tolos-2004}
L. Tolos, J. Schaffner-Bielich and A. Mishra, Phys. Rev. {\bf C 70} (2004) 025203.

\bibitem{Lutz:2009ff}
  M.~F.~M.~Lutz {\it et al.}  [PANDA Collaboration],
  arXiv:0903.3905 [hep-ex].

\bibitem{Lutz:2006ya}
  M.~F.~M.~Lutz and J.~Hofmann,
  Int.\ J.\ Mod.\ Phys.\ A {\bf 21}, 5496 (2006).

\bibitem{Wu:2010jy}
  J.~J.~Wu, R.~Molina, E.~Oset and B.~S.~Zou,
  Phys.\ Rev.\ Lett.\  {\bf 105}, 232001 (2010).

\bibitem{Wu:2010vk}
  J.~J.~Wu, R.~Molina, E.~Oset and B.~S.~Zou,
  Phys.\ Rev.\ C {\bf 84}, 015202 (2011).

\bibitem{Garcia-Recio:2013gaa}
  C.~Garcia-Recio, J.~Nieves, O.~Romanets, L.~L.~Salcedo and L.~Tolos,
  Phys.\ Rev.\ D {\bf 87}, 074034 (2013).

\bibitem{Xiao:2013yca}
  C.~W.~Xiao, J.~Nieves and E.~Oset,
  Phys.\ Rev.\ D {\bf 88}, 056012 (2013).

\bibitem{Isgur:1989vq}
  N.~Isgur and M.~B.~Wise,
  Phys.\ Lett.\ B {\bf 232}, 113 (1989).

\bibitem{Georgi:1990cx}
  H.~Georgi,
  Nucl.\ Phys.\ B {\bf 348}, 293 (1991).

\bibitem{Flynn:1992fm}
  J.~M.~Flynn and N.~Isgur,
  J.\ Phys.\ G {\bf 18}, 1627 (1992).

\bibitem{'tHooft:1973jz}
  G.~'t Hooft,
  Nucl.\ Phys.\ B {\bf 72}, 461 (1974).


\bibitem{Witten:1979kh}
  E.~Witten,
  Nucl.\ Phys.\ B {\bf 160}, 57 (1979).


\bibitem{Jenkins:1998wy}
  E.~E.~Jenkins,
  Ann.\ Rev.\ Nucl.\ Part.\ Sci.\  {\bf 48}, 81 (1998)

\bibitem{Matagne:2014lla}
  N.~Matagne and F.~Stancu,
  Rev.\ Mod.\ Phys.\  {\bf 87}, 211 (2015)

\bibitem{Gervais:1983wq}
  J.~L.~Gervais and B.~Sakita,
  Phys.\ Rev.\ Lett.\  {\bf 52}, 87 (1984).

\bibitem{Dashen:1993as}
  R.~F.~Dashen and A.~V.~Manohar,
  Phys.\ Lett.\ B {\bf 315}, 425 (1993)

\bibitem{Lee:1998pq}
  J.~P.~Lee, C.~Liu and H.~S.~Song,
  Phys.\ Rev.\ D {\bf 59}, 034002 (1999)

\bibitem{Chow:1999hm}
  C.~K.~Chow and T.~D.~Cohen,
  Phys.\ Rev.\ Lett.\  {\bf 84}, 5474 (2000)

\bibitem{Lee:2000wb}
  J.~P.~Lee, C.~Liu and H.~S.~Song,
  Phys.\ Rev.\ D {\bf 62}, 096001 (2000)

\bibitem{AzizaBaccouche:2001pu}
  Z.~Aziza Baccouche, C.~K.~Chow, T.~D.~Cohen and B.~A.~Gelman,
  Phys.\ Lett.\ B {\bf 514}, 346 (2001)

\bibitem{Wessling:2004ag}
  M.~E.~Wessling,
  Phys.\ Lett.\ B {\bf 603}, 152 (2004)
  [Phys.\ Lett.\ B {\bf 618}, 269 (2005)]

\bibitem{Cohen:2005bx}
  T.~D.~Cohen, P.~M.~Hohler and R.~F.~Lebed,
  Phys.\ Rev.\ D {\bf 72}, 074010 (2005)

\bibitem{Semay:2008wn}
  C.~Semay, F.~Buisseret and F.~Stancu,
  Phys.\ Rev.\ D {\bf 78}, 076003 (2008)


\bibitem{Lutz:2011fe}
  M.~F.~M.~Lutz, D.~Samart and A.~Semke,
  Phys.\ Rev.\ D {\bf 84}, 096015 (2011).

\bibitem{Lutz:2014jja}
  M.~F.~M.~Lutz, D.~Samart and Y.~Yan,
  Phys.\ Rev.\ D {\bf 90}, 056006 (2014).

\bibitem{Georgi:1990um}
  H.~Georgi,
  Phys.\ Lett.\ B {\bf 240}, 447 (1990).

\bibitem{Wise:1992hn}
  M.~B.~Wise,
  Phys.\ Rev.\ D {\bf 45}, 2188 (1992).

\bibitem{Casalbuoni:1996pg}
  R.~Casalbuoni, A.~Deandrea, N.~Di Bartolomeo, R.~Gatto, F.~Feruglio and G.~Nardulli,
  Phys.\ Rept.\  {\bf 281}, 145 (1997)

\bibitem{Mehen:2004uj}
  T.~Mehen and R.~P.~Springer,
  Phys.\ Rev.\ D {\bf 70}, 074014 (2004)


\bibitem{Yan:1992gz}
  T.~M.~Yan, H.~Y.~Cheng, C.~Y.~Cheung, G.~L.~Lin, Y.~C.~Lin and H.~L.~Yu,
  Phys.\ Rev.\ D {\bf 46}, 1148 (1992)
  [Phys.\ Rev.\ D {\bf 55}, 5851 (1997)].

\bibitem{Cho:1992gg}
  P.~L.~Cho,
  Phys.\ Lett.\ B {\bf 285}, 145 (1992).

\bibitem{Krause:1990xc}
  A.~Krause,
  Helv.\ Phys.\ Acta {\bf 63}, 3 (1990).

\bibitem{Lutz:2010se}
  M.~F.~M.~Lutz and A.~Semke,
  Phys.\ Rev.\ D {\bf 83}, 034008 (2011)

\bibitem{Luty:1993fu}
  M.~A.~Luty and J.~March-Russell,
  Nucl.\ Phys.\ B {\bf 426}, 71 (1994).

\bibitem{Dashen:1993jt}
  R.~F.~Dashen, E.~E.~Jenkins and A.~V.~Manohar,
  Phys.\ Rev.\ D {\bf 49}, 4713 (1994)
  [Phys.\ Rev.\ D {\bf 51}, 2489 (1995)].

\bibitem{Dashen:1994qi}
  R.~F.~Dashen, E.~E.~Jenkins and A.~V.~Manohar,
  Phys.\ Rev.\ D {\bf 51}, 3697 (1995).

\bibitem{Jenkins:1996de}
  E.~E.~Jenkins,
  Phys.\ Rev.\ D {\bf 54}, 4515 (1996).


%
%



%
%
\bibitem{GL84}
J. Gasser and H. Leutwyler, Ann. Phys. {\bf 158} (1984) 142.
%
%

\bibitem{Ecker89}
G. Ecker, J. Gasser, H. Leutwyler, A. Pich, and
E. De Rafael, Phys. Lett. B {\bf  223}  (1989) 425.


\bibitem{Borasoy}
B. Borasoy and U.-G. Mei\ss ner, Int. J. Mod. Phys. A {\bf 11} (1996) 5183.


\bibitem{Birse}
M. C. Birse, Z. Phys. A {\bf 355} (1996) 231.

\bibitem{Becher} T. Becher and H. Leutwyler, Eur. Phys. J. {\bf C 9} (1999) 643.

\bibitem{Fuchs}
T. Fuchs, M. R. Schindler, J. Gegelia and S. Scherer, Phys. Lett. B {\bf 575} (2003) 11.

\bibitem{Aaij:2015tga}
  R.~Aaij {\it et al.} [LHCb Collaboration],
  Phys.\ Rev.\ Lett.\  {\bf 115}, 072001 (2015)

\bibitem{Zhang:2009gy}
  Y.~J.~Zhang and Q.~Zhao,
  Phys.\ Rev.\ D {\bf 81}, 034011 (2010).

\bibitem{Chen:2011xk}
  D.~Y.~Chen and X.~Liu,
  Phys.\ Rev.\ D {\bf 84}, 034032 (2011).

\bibitem{Achasov:2012ss}
  N.~N.~Achasov and G.~N.~Shestakov,
  Phys.\ Rev.\ D {\bf 86}, 114013 (2012).

\bibitem{Limphirat:2013jga}
  A.~Limphirat, W.~Sreethawong, K.~Khosonthongkee and Y.~Yan,
  Phys.\ Rev.\ D {\bf 89}, 054030 (2014).

\bibitem{Yan:2009vr}
  Y.~Yan, K.~Khosonthongkee, C.~Kobdaj and P.~Suebka,
  J.\ Phys.\ G {\bf 37}, 075007 (2010).

\bibitem{Limphirat:2010zz}
  A.~Limphirat, C.~Kobdaj, P.~Suebka and Y.~Yan,
  Phys.\ Rev.\ C {\bf 82}, 055201 (2010).



\end{thebibliography}
\end{document}